\documentstyle[12pt]{article}
\newcommand\exa{\nopagebreak \begin{flushleft}\smallskip
\nopagebreak
\begin{minipage}[t]{5cm}\sloppy}
\newcommand\exb{\end{minipage}\kern%
1cm\begin{minipage}[t]{9cm}\sloppy}
\newcommand\exc{\end{minipage}\kern -3cm \smallskip%
\end{flushleft}}
\begin{document}
\begin{flushright}
UUITP, 12/1993 \\ hep-ph@xxx/9305318
\end{flushright}
\vspace{0.5 cm}

\begin{center}
{\Large \bf Rigid String From QCD Lagrangian}\\
\vspace{1cm}
{\bf A.R.Zhitnitsky}
\footnote{e-mail address is zhitnita@grotte.teorfys.uu.se \\
The address after 1 September, 1993 : Department of
Physics, SMU,Dallas, TX 75275-0175,USA }\\
\vspace{1cm}
Department of Theoretical Physics,Uppsala University,
S-75238, Uppsala, Sweden \\ and\\ Institute of Nuclear
Physics, Academy of Sciences, 630090 Novosibirsk, Russia.
\end{center}
\vspace{1cm}
\begin{abstract}

Starting from the 4-dimensional gluodynamics, we discuss
the statistical ensemble of torons ( the toron is a
self-dual solution with fractional topological number; it
can be understood as a point defect) which interact
strongly. It is shown that the effective Lagrangian
describing this statistical ensemble possesses the fourth
derivative kinetic term and leads to the area law for the
Wilson loop.

Besides that we derive the effective string action
containing the rigid (extrinsic curvature) term as a
consequence of the fourth derivative action.
\end{abstract}
\vspace {6cm}
\begin{flushleft} UUITP 12/1993 \\
May 1993
\end{flushleft}

\newpage

{\bf 1.Introduction.}
\vspace{0.5cm}

It is generally believed that the large -distances behavior
of the confining phase of QCD is given by an effective
string theory,see recent reviews
\cite{Gross},\cite{Polch}, \cite{Dine}.
The main problem of this idea can be formulated as follows.
How to find the correct collective variables in terms of
the underlying field theory (QCD) which effectively
describe the string theory.

In some simple models (it is a very narrow family of all
field theories), one has a map from original field theory
to stringlike variables, and one can derive properties of
the resulting "string theory" from the field theory.  One
of the well studied system of such kind is $2+1$
dimensional Ising-like model described by the
action\footnote{ Here and what follows we are discussing
the Euclidean version of the models}
\begin{equation}
\label{int1}
S=\frac{1}{2}\int
d^3x[\partial_{\mu}\phi\partial_{\mu}\phi+\lambda(\phi^2-
\frac{m^2}{\lambda})^2]
\end{equation}
As is known this model possesses the Bloch wall solution
which is independent of two variables ($t$ and $y$ )and
well localized in the transverse direction $z$. Such
solution spontaneously break translation invariance in the
$D-2$ transverse directions. As a result, there are
Nambu-Goldstone massless excitations about such background,
even if the field theory has only massive excitations about
trivial background. So the leading term in the effective
field theory describing low energy phenomena must take the
form
\begin{equation}
\label{int2}
S_{string}(f^i)=const\int dtdy [(\partial_t f^i)^2
+(\partial_y f^i)^2]+..., ~~~~~~~i=1, 2 ,D-2
\end{equation}
where $S_{string}(f^i)$ is the action governing low energy
phenomena, and we have chosen the $t,y$ plane as the plane
of the worldsheet.

The derivation of the $S_{string}(f^i)$ is standard
\cite{Wal}, \cite{Die}, \cite{Lus}, \cite{Lin}, \cite{Per}.
The steps involved in going from a theory with soliton-like
solution of eq.(\ref{int1}) to the effective string action
(\ref{int2}) are: a) to rewrite a functional integral $\int
D\phi\exp{iS(\phi)}$ about a soliton-like background in
terms of modes $f^i$that correspond to fluctuations of the
string and heavy modes $h$ that are separated by a mass
gap; b)to integrate out these massive modes. Thus one has:
\begin{equation}
\label{int3}
\int D\phi e^{iS(\phi)}\arrowvert_{about~ a ~classical~solution}=
\int Df^iDh e^{iS(f^i,h)}=\int Df e^{iS_{string}(f^i)}
\end{equation}

It was shown for particular model (\ref{int1}) by
\cite{Die} that the long wavelength expansion (\ref{int2})
for this model up to $0(\partial^6)$ reproduces the
Nambu-Goto action for the string
\begin{equation}
\label{int4}
S_{NG}\sim\sqrt{1+(\partial f)^2}=\sqrt{det h_{\mu\nu}}
\end{equation}
with induced metric
\begin{equation}
\label{int5}
h_{\mu\nu}=\delta_{\mu\nu}+\partial_{\mu}f\partial_{\nu}f.
\end{equation}
Besides that there are additional terms as well, which do
not have a geometrical interpretation. Few comments are in
order. First of all, the effective action (\ref{int2}) is
not renormalizable. This is not a catastrophe, however,
because the high frequency oscillations of $f$ are to be
cut off anyhow: when internal degrees of freedom of the
tube are excited, the description in terms of $f$ breaks
down.

As a second remark, let us note that the quantization of
the underlying field theory induces a quantization of the
effective string theory. As a consequence, Lorentz
invariance holds in the effective string theory for any $D$
(not only for $D=26$) \cite{Per}.

As a last remark we note that the absence of the so-called
rigid term
\cite{Pol}, \cite{Kle},
describing the extrinsic curvature of the world sheet, is
the direct consequence of the canonical expression for the
kinetic term (without higher derivatives) of the underlying
field theory (\ref{int1}). We'll come back to this point
later.

Thus the problem of derivation of the long distances
effective Lagrangian from field theory, which possesses the
soliton like solution reduces to the calculation of the
functional integral (\ref{int3}). The physical sense of the
string variable $f$ in this case is clear and can be
understood in terms of the original field variables $\phi$.
Indeed, roughly speaking, the string variable $f$ describes
the fluctuations about the classical solution $\phi_{cl}$,
see
\cite{Wal}-\cite{Per} for more details.

The gluodynamics {\em does not belong \/} to this class,
and so, the corresponding methods can not be applied to
Yang-Mills theory {\em directly}.  However, it is generally
believed that QCD might be represented as a string theory.
Many of the hadron properties will be understandable in
this case.  The standard approach to this problem is a
making of a guess what the effective long distances
lagrangian is going to be. But in such approach the
relation with the original QCD completely lost and the
connection of the string variables $f$ with gauge fields
$A_{\mu}^a$ looks absolutely unclear.

In this letter I shall follow in the opposite direction,
from the gauge fields to string theory. In this case, the
each step of the reformulation QCD in terms of string
variables is going to be under control (at least in
principle).  Besides that, the string variable $f$ can be
expressed in terms of the gauge theory.

Schematically, the steps involved in going from QCD to the
effective string theory look as follow.

\vspace{0.5cm}

\fbox{\rule[-0.3cm]{0.cm}{1cm}$S=-\frac{1}{4}\int
d^4x G_{\mu\nu}G_{\mu\nu} $}
\vspace{0.5cm}
$\stackrel{\Longrightarrow}{\bf{1}}$
\fbox{\rule[-0.3cm]{0.cm}{1cm}
$Z=\sum_{k=0}^{\infty} \frac{{\Lambda}^{4(k_1+k_2)}}%
{(k_1)!(k_2)!}
\sum_{I_{\alpha},q_{\alpha}}\prod_{i=1}^{k_1+k_2}d^4x_i
exp(-\epsilon_{int.})$ }
$\stackrel{\Longrightarrow}{\bf{2}} \\$
\vspace{0.5cm}

$\stackrel{\Longrightarrow}{\bf{2}}$
\fbox{\rule[-0.3cm]{0.cm}{1cm}$Z=\int D{\phi}
exp(-\int d^4x L_{eff.}(\phi))$}
$\stackrel{\Longrightarrow}{\bf{3}}$
\fbox{\rule[-0.3cm]{0.cm}{1cm}$
\langle W \rangle=
\int D f  exp(-\int d^2\sigma L_{string}(f))$}

Here the first step {\bf(1)} is related to the
consideration of the statistical ensemble of
pseudoparticles (point defects) with fractional topological
charge, so-called torons, which interact strongly. In the
next Section we briefly formulate the basic assumptions of
this toron approach. The second step{\bf(2)} is more or
less standard one which allows us to reformulate the
statistical mechanics problem in terms of the functional
integral over some auxiliary field $\phi$ with some
effective action $S_{eff}$. It turns out that this
$S_{eff}$ can be considered in the same way as $S$ from the
formula (\ref{int1}) in a sense that in both cases we have
some solitonic shape solution. So, the method briefly
described above for transition from $S$ (\ref{int1}) to
$S_{string}$ (\ref{int2}) by calculating of the functional
integral (\ref{int3}) can be applied in the case of the
gluodynamics as well. This is just step {\bf(3)}.

\vspace{0.5cm}

{\bf 2.Review of the toron approach}\footnote{See
\cite{Zhi1},\cite{Zhit}
for much of the material in this section.}

\vspace{0.5cm}

Before we proceed to the detail consideration of the string
representation of QCD, let me briefly formulate the basic
assumptions of the toron approach \footnote{ We keep the
term "toron", introduced in ref.\cite{Hoo1}. By this means
we emphasize the fact that the considering solution
minimizes the action and carries the topological charge
$Q=1/2$,i.e. it possesses all the characteristics ascribed
to the standard toron \cite {Hoo1}. However I should note
from the very beginning of this paper, that our solution
has nothing to do with the standard toron and it is
formulated in principle in another way than in ref.\cite
{Hoo1}. The keeping of this term has a historical origin.},
step {\bf(1)}.

i)I extend the class of admissible gauge transformation in
gluodynamics. Thus, I allow the configurations with
fractional topological charge (one half for $SU(2)$ group)
in the definition of the functional integral. It means that
a multivalued functions will appear in the functional
integral. However, the main physical requirement is - all
gauge invariant values must be singlevalued . Thus , the
different cuts accompany the multivalued functions should
be unobservable, i.e. the gauge invariant values coincide
on the upper and on the lover edges of the cut.

At large distances the {\em toron looks like a singular
gauge transformation \/}.  At small distances this
configuration should be somehow regularized.It can be
explicitly done for the separate toron, but a general
solution of this problem is still lacking. Fortunately, the
long distances pseudoparticle interaction (the expression
we are interested in) does not depend on regularization
procedure.

The direct consequence of the such definition of the
functional integral is the appearing of the new quantum
number classifying the vacuum states. Indeed, as soon as we
allowed one half topological charge,the number of the
classical vacuum states is increased by the same factor two
in comparison with a standard classification, counting only
integer winding numbers $|n>$.

Of course,vacuum transitions eliminate this degeneracy.
However the trace of enlargement number of the classical
vacuum states does not disappear. Vacuum states now
classified by two numbers : $0\leq \theta <2\pi$ and
$k=0,1$.  These is in agreement with large $N$ results
where the nontrivial $\theta$ dependence in pure YM theory
comes through $\theta/N$ \cite{Wit3} at large $N$ ( in
particular,$ <\tilde{G}G>\sim \sin(\frac{\theta}{N})$
\cite{Ven}).  Such a function can be periodic in $\theta$
with period $2\pi$ only if there are many vacuum states for
given values of $\theta$. These vacua should not be
degenerated due to the vacua transitions , however the
trace of the enlargement number of the vacuum states have
to be seen.

ii) The next main point of the toron approach may be
formulated as follows. We hope that in the functional
integral of the gluodynamics , when the bare charge tends
to zero and when we are calculating some long range
correlation function, only certain field configurations (
the toron of all types) are important. In this case the
hopeless problem of integration over all possible fields is
reduced to the problem of summation over classical toron
configurations. I have no proof that the system of
configurations which have been taken into account is a
complete system.  But I would like to stress that a lot of
problems ( like the $\theta$ dependence , the $U(1)$
problem , the counting of the discrete number of vacuum
states , the confinement , the nonzero value of the vacuum
energy and so on...) can be described in a very simple
manner from this uniform point of view.

Both these points are quite
nontrivial ones.  However , I would like to convince the
reader in the consistency of these assumptions by
considering a few simple models , where, from the one hand,
the results are well known beforehand and , from the other
hand they can be reproduced by toron calculations
\cite{Zhi1}.

Let me start by giving a few formulae from ref.\cite{Zhit}.
The grand partition function which presumably describes $4$
dimensional gluodynamics is given by
\begin{eqnarray}
\label{6}
Z=\sum_{k=0}^{\infty}
\frac{{\Lambda}^{4(k_1+k_2)}}{(k_1)!(k_2)!}
\sum_{I_{\alpha},q_{\alpha}}\prod_{i=1}^{k_1+k_2}d^4x_i
exp(-\epsilon_{int.}) ,~~~~~~~~~~~~        \\
\epsilon_{int.}=-\frac{4}{3}\sum_{i>j}q_i I_{i}
ln(x_i-x_j)^2
q_j I_{j} , \ \
\Lambda^{4-1/3} =c \frac{M_0^{4-1/3}}{g^2(M_0)}
\exp(-\frac{4{\pi}^2}{g^2(M_0)}) .  \nonumber
\end{eqnarray}
where two different kinds of torons classified by the
weight $I_i$ of fundamental representation of the $SU(2)$
group and $q_i$ is the sign of the topological charge.
Besides that, in formula (\ref{6}) the value $g^2(M_0)$ is
the bare coupling constant and $M_0$ is ultraviolet
regularization, so that eq.(\ref{6}) depends on the
renormalization invariant combination $\Lambda$.  In
obtaining (\ref{6}) we took into account that the classical
contribution to $Z$ from $k$ torons is equal to
\begin{equation}
\label{7a}
Z\sim \exp (-\frac{4{\pi}^2}{g^2}k).
\end{equation}
Besides that the factor $d^4x_i$ in eq.(\ref{6}) is due to
the 4 translation coordinates accompany an each toron and
combinatorial factor $k_1!k_2!$ is necessary for avoiding
double counting for $k_1$ torons and $k_2$ antitorons;
lastly, the average overall configurations ${q},{I}$ is an
average over all isotopical directions and topological
charge signs of torons.

To compute some vacuum expectation values it is convenient
to use the correspondence between the grand partition
function for the gas (\ref{6}) and field theory with
Sine-Gordon interaction , as it was done by Polyakov in
ref.\cite{Pol2} for 3d QED. Let us rewrite (\ref{6}) in the
form:
\begin{eqnarray}
\label{7}
Z_{\theta}=\int D\vec{{\phi}} exp(-\int d^4x L_{eff.})
,~~~~~~~~~~~~~~ \Box\equiv\partial_{\mu}\partial_{\mu},\\
L_{eff}=1/2(\Box \vec{\phi})^2-\sum_{\vec{I_{\alpha}}}
{\Lambda}^4\exp(i8\pi/\sqrt{3}\vec{I_{\alpha}}\vec{\phi}%
+i\theta/2)
-\sum_{\vec{I_{\alpha}}}{\Lambda}^4\exp
(-i8\pi/\sqrt{3}\vec{I_{\alpha}}\vec{\phi}-i\theta/2).
\nonumber
\end{eqnarray}
In this derivation it was used the fact that the logarithm
function which appears in the formula for the interaction
(\ref{6}) is the Green function for the operator
$\Box\Box$. After that we can use the method
\cite{Pol2}
to express the generating functional in terms of effective
field theory (\ref{7})\footnote{It is quite obviously,that
this Lagrangian does not correspond to any fundamental
theory.  In particular, the kinetic term has a forth
derivative form, so this theory does not describe any
asymptotic states in Minkowski space (there is no
continuation from Euclidean space here).  Thus, Lagrangian
(\ref{7}) is understood as effective one, describing the
statistical ensemble of pseudoparticles.}.

In this effective field theory the sum over
$\vec{I_{\alpha}}$ runs over the $2$ weights of the
fundamental representation of $SU(2)$ group. Note, that the
first interaction term is related to torons and the second
one to antitorons. Besides that, since we wish to discuss
the $\theta$ dependence , we also include a term
proportional to the topological charge density
$\frac{\theta}{32{\pi}^2}G_{\mu\nu}
\tilde{G_{\mu\nu}}$ to the starting Lagrangian
 and corresponding track from this to the effective
Lagrangian(\ref{7}).

{}From this expression it is clear that one of the feature of
the effective Lagrangian (\ref{7}) is the nontrivial
dependence on $\theta$ of the topological density and
susceptibility, relevant quantities for the solution of the
$U(1)$ problem :
\begin{equation}
\label{8}
 \frac{\delta
Z_{\theta}}{\delta{\theta}}\sim\langle\frac{1}%
{32{\pi}^2}G_{\mu\nu}
\tilde{G_{\mu\nu}}\rangle =i2{\Lambda}^4\sin(\theta/2),
-\pi\leq\theta\leq\pi.
\end{equation}

As discussed above, such a dependence on $\theta$ is in
agreement with large $N$ results. We mentioned here these
few consequences of the effective Lagrangian (\ref{7}) just
to demonstrate that the system (\ref{6}) reproduces these
very nontrivial, but well established at large $N$ results
( the correct $\theta/N $dependence, the number of vacuum
states equals $N$, the nonzero value for topological
density $\langle G_{\mu\nu}
\tilde{G_{\mu\nu}}\rangle \sim\sin(\theta/N)$ and so on)
correctly.
{}From the other hand, we expect ( see\cite{Zhi3} and
references therein) that all problems under consideration
(as well as confinement and string reformulation problems)
are tightly connected. Thus , {\em any self-consistent
dynamical solution of one of them should be necessarily
accompanied by the resolution of the rest problems within
same approach \/}.

Therefore we expect that the information about confinement
and string representation of QCD somehow is coded in the
effective Lagrangian (\ref{7}).

\vspace{0.5cm}

{\bf 3. String description of the gluodynamics.}

\vspace{0.5cm}

Now we want to discuss the relation between string variable
$f^i$ and auxiliary field $\phi$ from the effective
Lagrangian (\ref{7}).  But before to do so, we would like
to understand the physical sense of the $\phi$ field in
terms of the underlying gauge theory. To this aim we define
the $M$ as an operator that acts on original $A_{\mu}^a$
fields by gauge transforming them by $U^{x_0}(x)$; this
gauge transformation is singular at $x_0$ and has the
following property: for any plane crossing $x_0$ and for
any $x$ at the plane, as soon as $x$ encircles $x_0$, $U$
does not return to its original value ( as it happens in
the instanton case), but acquires a $Z_N $ phase ($N=2$ for
$SU(2)$ group):
\begin{equation}
\label{15b}
U^{x_0}(\alpha=2\pi)=\exp(-i2\pi/N)U^{x_0}(\alpha=0),\ \ \
U=\exp(iI\alpha)
\end{equation}
where $\alpha$ is an angle variable in the chosen $x-y$
plane and the point $x_0$ lies at the same plane.  {}From
its definition it must be clear that $M(x)$ absorbs one
half topological unit, so we say that $M(x)$ is the
annihilation operator for one point toron at $x$ with
weight $I$ and $M^+(x)$ is the creation operator for one
toron. It should be clear , that $U$ depends on all
$x_{\mu}$ variables , so that $M$ is the annihilation
operator for the point defect. However, at the chosen $x-y$
plane, $U$ depends only on angle variable $\alpha$.  The
singularity of $A_{\mu}=iU^+\partial_{\mu}U$ must be
smeared over an infinitesimal region around $x$ as it was
done for the separate toron solution, but the
regularization problem does not influent on the following
consideration.

We would like to express the $M$ in terms of the effective
field theory (\ref{7}).  To this aim, let us consider the
$\epsilon_{int}$ in the formula (\ref{6}) after the action
of the gauge transformation (\ref{15b}) at point $x_0$.
Because this gauge transformation creates an additional
toron at point $x_0$ with isospin $I_0$ in the system of
the other torons placed at $x_i$ with isospins $I_i$ we
will obtain an additional contribution to the
$\epsilon_{int}$. Namely, after action of the operator $M$
we have an additional interaction term between created
toron $I_0$ and torons $I_i$ from the system
\begin{equation}
\label{16b}
\Delta\epsilon_{int}\sim\sum_{i}I_0I_i\ln(x_0-x_i)^2.
\end{equation}
It is easy to understand that this interaction after simply
repeating the derivation of eq.(\ref{7}),reduces to the
following expression in the effective field theory:
\begin{equation}
\label{17b}
\langle M(x_0)\rangle=\int D\vec{{\phi}} exp(-\int d^4x L_{eff.})
\exp(i8\pi/\sqrt{3}\vec{I_0}\vec{\phi(x_0)}+i\theta/2)
\end{equation}
Thus, the operator $M$ under consideration in the effective
theory looks like this
\begin{equation}
\label{18b}
M_{\alpha}(x)=\exp(i\chi_{\alpha}(x)+i\theta/2),~~~~~~~~~~~~~
\chi_{\alpha}\equiv\frac{8\pi}{\sqrt{3}}\vec{I_{\alpha}}\vec{\phi}.
\end{equation}
Thus , the operator of large gauge transformation, $M$,
which should be highly non-local and nontrivial in terms of
the original fields ($A_{\mu}^a$ - gluons) has a very
simple form in terms of the auxiliary variables
$\chi_{\alpha}$.  It gives the link
$A_{\mu}\Longrightarrow\chi_{\alpha}$. From the other hand,
we shall see in a few moments that the $\chi_{\alpha}$
field related in a very simple manner to string variable
$f_i$. It gives the second wanted link
$A_{\mu}^a\Longrightarrow\chi_{\alpha}\Longrightarrow f_i$.

It is interesting to note that the disorder operator $M$
(\ref{18b}) in gluodynamics has the same exponential form
like in $2+1$ dimensional QED \cite{Snyd}.

Our next step is to consider the vacuum expectation value
of the Wilson loop.  As was explained above, the torons at
large distances look like singular pure gauge field with
definite isotopical direction and so, the $A_{\mu}^a$ field
is abelian at large distances (in a more detail see
\cite{Zhit}). Thus, the standard quasiclassical
approximation, when we substitute for $A_{\mu}^a$ the
corresponding classical solution, leads ( after simply
repeating the derivation of (\ref{7})) to the following
expression for $<W>$ at $\theta=0$
\begin{eqnarray}
\label{25b}
\langle W \rangle=
\langle Tr\exp(\oint_{l}iqA_{\mu}dx_{\mu})
\rangle =~~~~~~~~~~~~~~~~~\\
\int D\chi_{\alpha}  exp(-\int d^4x L_{eff.}),\ \
L_{eff}=1/2(\frac{\sqrt{3}}{4\pi})^2 (\Box
\chi_{\alpha})^2-\sum_{\vec{I_{\alpha}}}
2{\Lambda}^4\cos(\chi_{\alpha}+
\vec{I_{\alpha}}\vec{\Phi}) .    \nonumber
\end{eqnarray}
where the term proportional to $\Phi$ is related to Wilson
loop insertion and has the following property : $\Phi(x)$
is equal to the external charge $2\pi q$ if $x\in S$,
Wilson plane, and $\Phi=0$ otherwise.  In this derivation
we took into account that if the toron is in the $S$ plane,
then the integral over $G_{\mu\nu}d\sigma_{\mu\nu}$ is
non-zero, and it is equal to zero otherwise (see for a more
detail about toron properties the ref .\cite{Zhit} ).

At this moment I would like to come back to discussion in
the Introduction concerning of the semiclassical
calculation of the functional integral in the solitonic
background, see formula (\ref{int3}). For this purpose we
will be considering the effective Lagrangian from the
formula (\ref{25b}) on the same foot as fundamental
Lagrangian from the formula (\ref{int1}).  To this aim, we
have to look for the solution of the corresponding
classical equation :
\begin{equation}
\label{26b}
\Box\Box{\chi}_{cl}^{\prime}+4{\Lambda}^4
(\frac{4\pi}{\sqrt{3}})^2\sin
({\chi}_{cl}^{\prime})
=2\pi\theta_S(z,t){\delta}'(x){\delta}'(y),
\end{equation}
where ${\chi}^{\prime}\equiv\chi+\vec{\Phi}\vec{I}$, $
{\delta}'(x)
\equiv\frac{d\delta(x)}{dx}$
,$\theta_S(z,t)=1$ if $z,t \in S$ and $\theta_S(z,t)=0$
otherwise.  The right- hand side of this equation is
related to the Wilson loop insertion,i.e. with function
$\Phi(x)$ \footnote{ More exactly, the right hand side is
proportional to $2
{\delta}'(x){\delta}'(y)+{\delta}'''(x)sign
(y)\delta_{y,0}+ {\delta}'''(y)sign (x)\delta_{x,0}$ where
$\delta_{x,0}$ is Kroneker symbol. However the last two
terms do not play any role in the following calculation and
we will skip them for the simplification of formulae. The
technical reason for that is the vanishing of the integrals
like $\int dy\delta_{y,0}G(y)=0$ for any smooth function
$G(y)$.}.

I do not know an exact solution for this equation, but the
physics suggests that a linearization is legitimate, so for
qualitative estimation we can substitute $\chi$ instead of
$\sin(\chi)$ and find $\chi_{cl}$ by means of Fourier
transformation ( here and what follows we drop the prime in
the notation for $\chi '$ field )
\begin{equation}
\label{st1}
\chi_{cl}\sim\theta_S(z,t)\int d^2k e^{ikx}k_x
k_y\frac{1}{k^4+m^4}, ~~~~~~~~~~m^4\equiv 4{\Lambda}^4
(\frac{4\pi}{\sqrt{3}})^2
\end{equation}

This solution correctly reproduces the discontinuity
related to the right hand side of eq.(\ref{26b}). Besides
that, the integral (\ref{st1}) can be reduced to the
modified Bessel function $K_0 (z)$ and so we have the
exponentially localized in the transverse directions
$(x,y)$ solution, more exactly
\begin{equation}
\label{st2}
\chi_{cl}(|x|\rightarrow\infty)\sim
i\partial_{x}\partial_{y}\lgroup
\exp(-e^{\frac{i\pi}{4}}m|x|)-
\exp(-e^{-\frac{i\pi}{4}}m|x|)\rgroup, ~~~~~~~~~
|x|\equiv\sqrt{x^2+y^2}
\end{equation}

Thus, this situation remind us the analysis of the $2+1$
dimensional model (\ref{int1}) with the well-localized
Bloch wall solution. Now we can expand the effective action
(\ref{25b}) in the background of the classical solution
(\ref{st1}) as it was done for $2+1$ dimensional model
(\ref{int3}), integrate over perpendicular $x,y$ directions
and end up with some effective 'string' theory \footnote{It
is clear, that we have essentially a nontrivial dependence
only on two variables, $x,y$. The shape (\ref{st2})
guarantees the convergence of the corresponding integrals
$\int dxdy$, over the direction perpendicular to the Wilson
loop. So, starting from 4 dimensional action we end up with
2 dimensional 'string' theory.}.

But before to do so, I would like to make a few comments.
First of all, there is a big difference between the induced
string action derived from the underlying field theory
(\ref{int1}) and from the effective Lagrangian (\ref{25b}).
In the former case we have the classical solitonic solution
corresponding to domain wall with
$\phi(+\infty)\neq\phi(-\infty)$.  There are no sources
(quarks) at all. So, we have infinitely long string.  In
the later case we are calculating the vacuum expectation
value of the Wilson loop, so we have inserted the heavy
quarks into the system. It means that we are describing an
open string with the fixed ends.

As a consequence of the Wilson loop insertion, the right
hand side of the eq.(\ref{26b}) has a very important
significance: the classical solution (\ref{st1}) is
appearing together with the sources. It means that the
string will emerge into the theory simultaneously with
source insertion.

The same situation takes place in 3-dimensional QED
\cite{Pol2}, but in this case the equation analogous
(\ref{26b}) is essentially one dimensional. Right hand side
of the corresponding equation is proportional to
$2\pi\theta_S(z,t){\delta}'(x)$ as a consequences of the
Wilson loop insertion. Such nonlinearity ensures the
exponentially localized in $x$ direction solution and
reproduces the correct discontinuity of $\chi_{cl}$
solution. As a consequence of it we have an area law in the
model.

Now we are in position to describe the low energy effective
action.  Because our solution (\ref{st1}) spontaneously
breaks translation invariance in the $D-2=2$ transverse
space dimensions, there are Nambu-Goldstone massless
excitations about such background. The derivation of the
corresponding $S_{string}$ is standard and shortly
discussed in the Introduction. However, for our purpose it
is enough to keep only a few leading terms in the low
energy expansion, so we can write the fluctuations of the
$\chi$ field in the following form
\begin{equation}
\label{st3}
\chi (t,z,x_i)=\chi_{cl}\lgroup
x'_i=x_i+f_i(t,z)\rgroup +0(f^2),~~~~~~~i=1,2
\end{equation}
where vector field $f^i$ can be treated as a string
variable and represents the deflection of the thin flux
tube from its rest position \footnote{Actually the
thickness of the string is order $\Lambda$, the only
dimensional parameter we have in gluodynamics. For
justification of procedure, using in the text, see
discussion at the end of this Section}.  Using the
decomposition (\ref{st3}) the $(\Box\chi)^2$ can be
represented in the following form
\begin{eqnarray}
\label{st4}
(\Box\chi)^2=\lgroup\frac{{\partial}^2\chi_{cl}}{\partial
x_k^2}\rgroup^2
+2\lgroup\frac{{\partial}^2\chi_{cl}}{\partial
x_k^2}\rgroup
\lgroup\frac{{\partial}^2\chi_{cl}}%
{\partial x_i\partial x_j}
\frac{\partial f_i}{\partial x_{\mu}}%
\frac{\partial f_j}{\partial x_{\mu}}
+\frac{\partial\chi_{cl}}{\partial x_i}%
\frac{{\partial}^2 f_i}{\partial x_{\mu}
\partial x_{\mu}}\rgroup+
\\ \nonumber
\lgroup\frac{{\partial}^2\chi_{cl}}{\partial x_i%
\partial x_j}
\frac{\partial f_i}{\partial x_{\mu}}%
\frac{\partial f_j}{\partial x_{\mu}}
+\frac{\partial\chi_{cl}}{\partial x_i}%
\frac{{\partial}^2
f_i}{\partial x_{\mu}
\partial x_{\mu}}\rgroup^2
\end{eqnarray}
Here $x_i,~~i=1,2$ are variables, perpendicular to Wilson
loop. They describe the classical solution (\ref{26b}). At
the same time $x_{\mu},~~\mu=0,3$ variables describe the
string plane. The position of the string is specified by a
two-component vector field $f_i$, depending on $x_{\mu}$.

If we now insert this configuration (\ref{st3}) into the
functional integral (\ref{25b}) and perform the $x_i$
integrations we shall obtain some effective string
Lagrangian which, from the very general arguments, must be
local and invariant under the Poincare transformations in
$x_{\mu}\equiv(z,t)$ plane. Besides that this Lagrangian
must be invariant under $0(2)$ rotations and translations
of the vector field $f_i$ \cite{Lus}.  Of course, all these
properties will be fulfilled automatically in our scheme.
Explicit integration over $d^2 x_i $ gives the following
formula for effective string action describing the long
wavelength fluctuations of the string (we are keeping only
leading terms of the expansion, proportional to $(f_i)^2$):
\begin{equation}
\label{st5}
S_{string}=\Lambda^2\int d^2 x_{\mu}\{c_1+c_2\lgroup
\frac{\partial f_i}{\partial x_{\mu}}%
\frac{\partial f_i}{\partial x_{\mu}}%
\rgroup+c_3{\Lambda}^{-2} \lgroup%
\frac{{\partial}^2 f_i}{\partial x_{\mu}
\partial x_{\mu}}\frac{{\partial}^2 f_i}%
{\partial x_{\nu}
\partial x_{\nu}} \rgroup+0(f^4)+...\}.
\end{equation}
Here, $c_1,c_2,c_3$ are dimensionless constants determined
by classical solution $\chi_{cl} (x_i)$. In particular, the
constant $c_1$ is two-dimensional classical action:
\begin{equation}
\label{st6}
c_1={\Lambda}^{-2}\int d^2 x_i L_{eff}(x_{cl}),~~~
L_{eff}=(\frac{\sqrt{3}}{4\pi})^2
(\chi_{cl}-\eta)\Box\Box(\chi_{cl}-\eta)-4%
\Lambda^4[\cos(\chi_{cl})-1]
\end{equation}
and $c_2$ and $c_3$ are defined by the following integrals
\begin{eqnarray}
\label{st7}
c_2=\Lambda^{-2}(\frac{\sqrt{3}}{4\pi})^2
\int d^2x_i(\chi_{cl}-\eta)\Box\Box(\chi_{cl}-\eta) \\
c_3=(\frac{\sqrt{3}}{4\pi})^2
\int d^2x_i(\chi_{cl}-\eta)\Box(\chi_{cl}-\eta).\nonumber
\end{eqnarray}
Here $\eta$ is related to Wilson loop insertion and has the
property that $\Box\Box\eta\sim{\delta}'(x){\delta}'(y)$
and $\eta\sim sign (y)\delta_{y,0} sign(x)\delta_{x,0}$,
see footnote after formula (\ref{26b}).  Now we can
estimate these coefficients using the approximate
expression for the $\chi_{cl}$ from (\ref{st1}). We expect
that the accuracy for such procedure is not very high and
the numerical coefficient given bellow should be considered
as an estimation of the order of value. With these remarks
in mind we obtain:
\begin{equation}
\label{st8}
c_1\simeq
2c_2\simeq\frac{\sqrt{3}\pi}%
{32},~~~~~~c_3\simeq\frac{3}{256\pi}.
\end{equation}
We emphasize that the convergent result for these
coefficients is the direct consequence of the fourth
derivative term in the action as well as correct magnitude
for the discontinuity related to Wilson insertion.
Technically it can be seen from the following expression
for the $\Box\Box(\chi_{cl}-\eta)$:
\begin{equation}
\label{st9}
\Box\Box(\chi_{cl}-\eta)(x)\sim
\int d^2k e^{ikx}k_x k_y\lgroup\frac{k^4}{k^4+m^4}-1\rgroup\sim
\int d^2k e^{ikx}k_x k_y\lgroup\frac{m^4}{k^4+m^4}\rgroup.
\end{equation}

On substitution of (\ref{st9}) into the
formulae(\ref{st6},\ref{st7}) we obtain the convergent
integral at large $k$ ($x,y\rightarrow 0$) as was announced
\footnote{The calculation of $c_3$ can be easily done by
means of Fourier transformation, using the identity
$\Box={\Box}^{-1}\Box\Box$ and acting by operator
$\Box\Box$ on $(\chi_{cl}-\eta)(x)$ as shown in (\ref{st9})
and substituting instead of ${\Box}^{-1}$ its Fourier
transformed expression $k^{-2}$.}.

Now we want to rewrite the effective string action
(\ref{st5}) in geometrical terms. With this aim we note
that the first two terms in the derivative expansion can be
represented in the Nambu-Goto form:
\begin{equation}
\label{st10}
S_{NG}=\frac{1}{2\pi\alpha'}\int d^2 x_{\mu}%
\sqrt{1+\lgroup
\frac{\partial f_i}{\partial x_{\mu}}%
\frac{\partial f_i}{\partial x_{\mu}}%
\rgroup}=\frac{1}{2\pi\alpha'}\int d^2 x_{\mu}\sqrt{det h}
\end{equation}
with induced metric
\begin{equation}
\label{st11}
h_{\mu\nu}=\delta_{\mu\nu}+\frac{\partial f_i}{\partial
x_{\mu}}\frac{\partial f_i}{\partial
x_{\mu}},~~~~~~~h\equiv det h_{\mu\nu}
\end{equation}
 and string tension $(2\pi\alpha')^{-1}\equiv
c_1{\Lambda}^2$.  The third term of the expansion
(\ref{st5}) can be rewritten (up to higher corrections) as
the extrinsic curvature
\begin{equation}
\label{st12}
S_{K}=\frac{1}{{\alpha}_r}\int d^2 x_{\mu}%
\sqrt{h}K^{i}_{\mu\nu}K^{i\mu\nu}=
\frac{1}{{\alpha}_r}\int d^2 x_{\mu}\sqrt{h}%
\lgroup\Delta(h)f^i\rgroup^2,~~~~
\Delta(h)f^i=\frac{1}{\sqrt h}\partial_{\mu}
(\sqrt{h}h^{\mu\nu}\partial_{\nu}f^i),
\end{equation}
where $K^{i}_{\mu\nu}$ is known as the second fundamental
form and ${\alpha}_r^{-1}\equiv c_3$ is the rigidity
parameter.

Thus, our effective string action in the leading order in
$f$ can be written as follows
\begin{equation}
\label{st13}
S_{string}=S_{NG}+S_K+ ...
\end{equation}
This new, rigid term was introduced to string theory in
refs.\cite{Pol},\cite{Kle}.  It is easy to see that $S_K$
is the invariant under the scale transformation and
$\alpha_r$ is dimensionless constant. The motivation
\cite{Pol} for the inclusion of the extrinsic curvature
term to the string action is some desire to get a "smooth"
string. Indeed with only intrinsic terms, surfaces can
crumple up over arbitrary short distances, as long as their
total area is preserved. The extrinsic curvature acts to
give the surface rigidity, smoothing it out over short
distances. This property is quite desirable for QCD.

The formula (\ref{st13}) with calculable (in principle)
coefficients $\alpha' , \alpha_r$ is the main result of
this letter.  Now several comments are in order.

i)The quantization of the underlying field theory
(4-dimensional YM) induces a quantization of the induced
string theory (\ref{st13}).  In particularly, it is clear,
that the light cone quantization of the fundamental string
is not relevant to quantization of the effective string
(\ref{st13}). For instance, we would expect that the
Lorentz invariance of the effective string is a good
symmetry not only for dimension $D=26$, but for $D=4$ as
well. It has been checked explicitly for a more simple
$2+1$ dimensional model \cite{Per}, and I believe that the
same is true in our case as well.

ii) It is well-known that the higher derivative Lagrangian,
defined in Minkowski space -time, violates the unitarity
because of exponentially growing modes in time (see, e.g.
\cite{Polch},\cite{Bra}). However, a priori, there is
nothing wrong with theories of surfaces embedded in
Euclidean space -time and described by the extrinsic
curvature term. Anyhow, we are considering the Lagrangian
(\ref{25b}) with higher derivative terms, as an {\em
effective} one, describing our statistical ensemble of
pseudoparticles (\ref{6}). The appearance of the fourth
derivative term in this Lagrangian (as well as extrinsic
curvature term $S_K$ in the string effective action
(\ref{st12})) is the direct consequence of the strong
$\sim\ln(x_i-x_j)^2$ pseudoparticle interaction at large
distances.

iii) It is useful to treat our effective string Lagrangian
like the chiral Lagrangian (describing $\pi$-meson physics)
in a sense that only lightest degrees of freedom are
relevant to the problem.  In such treatment the
small-distance physics (regularization, renormalization,
loop calculation and so on...) is coded in the magnitude of
constants $c_i$.

iv) As a next remark, I would like to comment the result
\cite{Han}, concerning the rigid string. It was shown that
starting from renormalizable unitary field theory one can
get the higher derivative terms in the effective action by
integrating out heavy fields from underlying field theory.
However, in such procedure one gets the wrong sign for the
fourth-order derivative terms in effective action and the
corresponding strings are not smooth.

We would like to note, that the effective action
(\ref{25b}), we are dealing with, has different origin. It
describes the statistical ensemble (\ref{6}) of
pseudoparticle. The corresponding Partition Function is
well-defined; the field $\chi$ which appears in the
effective action is auxiliary one and was introduced just
to describe this statistical ensemble. It is clear that
$\chi$ field {\em does not} describe the asymptotic states
and thus, the argumentation of
\cite{Han} can not be applied
to this case. So, we should not be surprised

that the string action we derived (\ref{st13}) has the
correct sign for rigid term.  The technical reason for that
can be easily seen from the expression for $(\Box\chi)^2$,
(\ref{st4}). The relative positive sign in this
decomposition leads to the positive sign between Nambu-Goto
$S_{NG}$ and rigid $S_K$ terms in the string action
(\ref{st13}).

v) We keep only a few leading terms in the low energy
expansion.  To include higher order effects in $f$, one
needs to integrate over all massive excitations in the
classical background, as it was done for simple models in
\cite{Wal}-\cite{Per}. It is clear that in such procedure
we will get an infinite number of terms. Some of these
terms have a geometrical interpretation, some of them- not.
But we expect that in the long wavelength limit only a few
leading terms ( which have a very clear geometrical and
physical meaning) of this expansion are important and they
are given by formula (\ref{st13}).  Here some arguments in
favor of this hope.

But before to give these arguments, let us try to answer on
the following question. Whether the effective string
description, obtained from underlying field theory
describes the long wavelength limit correctly?  In other
words, is it possible to choose some parameters of the
original theory so that the energy of fluctuations to be
too small to excite the internal structure of the string.
In this case the string can be considered as the
structureless with zero width. Nielson and Olesen
\cite{Niel} addressed this question in the Abelian Higgs
model and they have shown that the string is effectively of
zero width when the length scale of the energy levels for
excitations of the string (defined by string tension
$\alpha'$ ) is much greater than the length scales
characterizing the width of the string (the penetration
depth and the correlation length in the Abelian Higgs
model).  In the model this requirement corresponds to the
electric charge much bigger than one, $e\gg 1$, thus the
thin string condition is the strong coupling limit,
$\hbar\rightarrow\infty$, which makes semiclassical
approximation very doubtful.

The same situation takes place for the model (\ref{int1}),
see \cite{Per}. Indeed in this case the classical string
solution in eq. (\ref{int1}) has a width of order $m^{-1}$;
the energy scale for excitations of the string is given by
\begin{equation}
\label{st14}
\frac{1}{2\pi\alpha'}\sim\int dz L_{cl}(z)%
\sim\frac{m^3}{\lambda}.
\end{equation}
The string can be considered as thin one when the internal
modes will not be excited, i.e.
\begin{equation}
\label{st15}
m^2\gg\frac{1}{2\pi\alpha'},~~~~ \Longrightarrow \lambda\gg m
\end{equation}
As before, this condition means the strong coupling limit
and it is not clear whether the standard semiclassical
approximation can be applied to this system.

In contrast with these explicit strings, I would expect
that QCD (implicit) string, we are interested in, has
different features. Let me demonstrate this point by
considering the relation analogous to (\ref{st15}). But
first of all I should note, that in contract with $2+1$
dimensional models discussed above, in gluodynamics we have
the only parameter in the theory, $\Lambda$, and thus , all
relations like (\ref{st15}) have numerical and not
parametrical meaning.  In our approximation the string
tension $(2\pi\alpha')^{-1}$ is equal to $ c_1{\Lambda}^2$.
At the same time, the characteristic scale of the internal
excitations of the string is determined by its width end is
equal to $m$ (\ref{st1},\ref{st2}).  The criterion for
string to be {\it thin} when fluctuations of the string
will not excite the internal modes, looks as follows
\begin{equation}
\label{st16}
m^2\gg\frac{1}{2\pi\alpha'},~~~~ \Longrightarrow%
2{\Lambda}^2 (\frac{4\pi}{\sqrt{3}})\gg c_1{\Lambda}^2
\end{equation}
With our very rough estimation for coefficient $c_1$
(\ref{st8}), the criterion (\ref{st16}) is satisfied.
Probably this numerical smallness for
$\frac{1}{2\pi\alpha'}$ is related, somehow, to $1/N$
expansion.  We are considering this numerical game as a
hint suggesting how the structureless string could appear.
In some sense, the inequality (\ref{st16}) is justification
for our expansion (\ref{st3}), see footnote after this
formula.

\vspace{0.5cm}

{\bf 4.Final remarks}.
\vspace{0.5cm}

The main point of this letter can be formulated as follows.
We believe that most of the fundamental problems in QCD,
such as the $\theta/N$ dependence, N vacuum states,
confinement, string representation of QCD, and so on,
should be solved at the same time within the same dynamical
approach.  Some of these problem can be understood within
so-called "toron approach".  In particular, from the
corresponding effective Lagrangian (\ref{7}) it is possible
to reproduce the correct behavior for the vacuum
expectation value of the topological density $
<\tilde{G}G>_k\sim \sin(\frac{\theta +2\pi k}{N})$ and
number of vacuum states $N$. Therefore, we would expect the
information about confinement (and string representation as
a consequence of it) is coded in the same Lagrangian.

We have demonstrated the possible way of extracting this
information from Lagrangian.  The result is the formula
(\ref{st13}). Let us note that the string fields $f_i$ in
the formula are not the space-time coordinates, but some
variables related to color space. This is hardly surprising
because the 't Hooft's analysis \cite{Hoof} of the large
$N$ behavior (this is the main motivation of our belief in
string picture of QCD ) ensures planarity in index space
and not in real space-time.  The main fundamental ( not
technical) assumptions I made in this derivation are
following :

i)The multivalued functions are admissible in the
definition of the functional integral. This assumption is
related to a new classification of the vacuum states.

ii) The only certain field configurations ( torons) are
important and the problem of integration over all possible
fields is reduced to the problem of summation over
classical toron configurations.

 It is quite possible that the technical realization of it
can be given in a different, more appropriate way than we
discussed. But I believe that the main point of this
Letter, the new classification of gauge fields, will emerge
in the formulation of the Theory.

\vspace{0.5cm}

{\bf Acknowledgments}
\vspace{0.5cm}

I am very grateful to G. Veneziano for valuable and
stimulating discussion.

This work was supported by the NFR grant F-FU 6821-301.

\end{document}